\documentclass[final,3p,times]{elsarticle}
\usepackage{amssymb}
\usepackage{hyperref}
\usepackage{graphicx}
\usepackage{multirow}
\usepackage{enumitem} 
\usepackage{amsmath}
\usepackage{siunitx}
\usepackage{float}
\usepackage{rotating}
\usepackage{xcolor}
\usepackage{natbib}
\biboptions{authoryear}

\definecolor{custompurple}{RGB}{128, 0, 128}

\journal{arXiv}

\begin{document}

\begin{frontmatter}

\title{The ULS23 Challenge: a Baseline Model and Benchmark Dataset for 3D Universal Lesion Segmentation in Computed Tomography}

\affiliation[inst1]{organization={Department of Radiology and Nuclear Medicine, Radboud University Medical Center},
            city={Nijmegen},
            country={The Netherlands}}
\affiliation[inst2]{organization={Department of Radiology, Jeroen Bosch Hospital},
            city={‘s-Hertogenbosch},
            country={The Netherlands}} 
\affiliation[inst3]{organization={Department of Radiology, University Medical Center Groningen},
            city={Groningen},
            country={The Netherlands}} 
\affiliation[inst4]{organization={Fraunhofer Institute for Digital Medicine MEVIS},
            city={Bremen},
            country={Germany}}

\author[inst1]{M.J.J. de Grauw}
\author[inst1]{E.Th. Scholten}
\author[inst1]{E.J. Smit}
\author[inst1,inst2]{M.J.C.M. Rutten}
\author[inst1,inst3]{M. Prokop}
\author[inst1,inst4]{B. van Ginneken}
\author[inst1,inst4]{A. Hering\corref{cor1}}
\ead{alessa.hering@radboudumc.nl}
\cortext[cor1]{Corresponding author}

\begin{abstract}
Size measurements of tumor manifestations on follow-up CT examinations are crucial for evaluating treatment outcomes in cancer patients. Efficient lesion segmentation can speed up these radiological workflows. While numerous benchmarks and challenges address lesion segmentation in specific organs like the liver, kidneys, and lungs, the larger variety of lesion types encountered in clinical practice demands a more universal approach. To address this gap, we introduced the ULS23 benchmark for 3D universal lesion segmentation in chest-abdomen-pelvis CT examinations. The ULS23 training dataset contains 38,693 lesions across this region, including challenging pancreatic, colon and bone lesions. For evaluation purposes, we curated a dataset comprising 775 lesions from 284 patients. Each of these lesions was identified as a target lesion in a clinical context, ensuring diversity and clinical relevance within this dataset. The ULS23 benchmark is publicly accessible at \url{https://uls23.grand-challenge.org}, enabling researchers worldwide to assess the performance of their segmentation methods. Furthermore, we have developed and publicly released our baseline semi-supervised 3D lesion segmentation model. This model achieved an average Dice coefficient of 0.703 ± 0.240 on the challenge test set. We invite ongoing submissions to advance the development of future ULS models.
\end{abstract}


\end{frontmatter}

\section{Introduction}
\label{sec:intro}
The volume of CT examinations conducted annually continues to rise~\citep{masjedi2020european}, leading to higher workloads for radiologists~\citep{mcdonald2015effects}. With the global cancer burden predicted to increase by 47\% in 2040 compared to 2020~\citep{sung2021global}, oncological radiology can be expected to significantly contribute to these increasing demands. Cancer patients frequently undergo several imaging examinations during their treatment and subsequent disease monitoring~\citep{rehani2020patients}. In addition, there is a growing interest on the early detection of cancer through imaging~\citep{crosby2022early,adams2023lung}. Computer-assistance during reading may help radiologists successfully manage this increasing workload. \\

With minimal guidance from radiologists, automatic segmentation models can reduce the time burden and inter-observer variability associated with manually annotating lesions in oncological scans. Lesions can be selected for segmentation using a single-click~\citep{Tang2020_OneClick}, bounding box annotation~\citep{mazurowski2023segment, ma2024segment} or via a detection model~\citep{Yan2019_MULAN}. The longitudinal measurements obtained from segmented lesions can then be analyzed according to clinical guidelines such as the Response Evaluation Criteria In Solid Tumors (RECIST)~\citep{Eisenhauer2009,schwartz2016recist}. Additionally, automatic 3D segmentation of lesions facilitates more sophisticated analyses, such as using radiomics features~\citep{gillies2016radiomics} to distinguish between lesion subtypes. Moreover, registration algorithms can be used to propagate segmented lesions to follow-up scans, which can lead to considerable time savings during subsequent examinations~\citep{hering2024improving}. \\

Over the past decade, AI-driven automatic tumor segmentation models have made considerable progress for certain high-profile lesion types, such as those found in the liver~\citep{bilic2019liver}, kidney~\citep{heller2023kits21}, or lungs~\citep{pedrosa2019lndb}. However, segmentation models for a wider range of lesion types, particularly within the often-examined thorax-abdominal region, remain relatively underexplored. The development of these Universal Lesion Segmentation (ULS) models, necessitates a diverse training dataset. Whilst there has been previous work on ULS (see section \ref{sec:rw_uls}), most studies have relied heavily on the DeepLesion dataset. This dataset only contains lesion diameter annotations on a single axial slice and is thus not particularly suited for developing 3D segmentation models. Furthermore, ground-truth segmentation masks used during evaluation on this dataset by previous publications~\citep{Cai2018_SlicePropogated, Tang2020_OneClick} are not publicly available, limiting reproducibility.  It is also uncommon for ULS models to be released to the public, hindering their integration into annotation workflows for researchers or for further clinical assessment.\\ 

\noindent \textbf{Contributions:} \\
In light of these conditions, we have launched the ULS23 challenge which,

\begin{itemize}
  \item Encourages advancements in model performance by collecting a large and diverse training dataset. We introduce two new datasets targeting lesions in the pancreas and bones, which are traditionally challenging to segment. Additionally, we aggregate 10 publicly available datasets with a lesion segmentation component into a single, easily accessible data repository.
  \item Enhances the reproducibility of ULS research by establishing a reliable benchmark using a carefully curated test set consisting of clinically relevant lesions from two Dutch medical centers.
  \item Facilitates the research community's access to state-of-the-art ULS models by developing and releasing our baseline semi-supervised ULS model to the public.
\end{itemize}

\section{Related Work}
\label{sec:rw}

\subsection{Related Biomedical Grand Challenges}
\label{sec:rw_mc}
Since the introduction of the SLIVER liver segmentation challenge, at the 2007 Medical Image Computing and Computer-Assisted Intervention (MICCAI) conference~\citep{heimann2009comparison}, biomedical grand challenges have become an integral part of the medical image analysis field. In these events, participants compete against each other within a limited timeframe to develop the best-performing solution for a given task, evaluated using a pre-determined set of metrics. This approach enables direct comparison of a wide variety of model architectures and training methods. Biomedical challenges often release novel annotated datasets to the public to promote open-source solutions and democratize model development. Several previous challenges have included lesion segmentation components, such as the LiTS challenge~\citep{bilic2019liver} for liver tumor segmentation, LNDb~\citep{pedrosa2019lndb} which incorporated a lung nodule segmentation subtask, and KiTS19~\citep{heller2019kits19} which required segmentation of kidney lesions.

\subsection{The nnUnet Framework}
\label{sec:rw_nnunet}
One influential method which has won multiple medical segmentation challenges is the \textit{no-new-Unet} (nnUnet) framework developed by \cite{isensee2021nnu}. It trains a basic convolutional encoder-decoder network that adaptively extracts model parameters from a task's dataset, which was demonstrated to be an effective driver of performance. Their results on the Medical Segmentation Decathlon, a challenge for comparing generalizable medical segmentation methods~\citep{antonelli2022medical}, showed that the nnU-Net compared favorably to the vast majority of more complex models at the time~\citep{isensee2019nnu}, emphasizing the importance of benchmarking architectural modifications against a well-tuned baseline on a carefully curated dataset. We have made extensive use of this framework to develop and benchmark our baseline method.

\subsection{DeepLesion}
\label{sec:rw_dl}
The first public dataset tailored specifically for universal lesion detection and segmentation was released in 2017~\citep{Yan2018_DeepLesion,Yan2018_DeepLesionGraphs}. It contains a total of 32,735 lesions from 4,427 unique patients across 10,594 studies. The annotations for the dataset were collected from the Picture Archiving and Communication Systems (PACS) of the National Institutes of Health Clinical Center. During clinical practice in these institutes, radiologists routinely used bi-directional measurements to determine the size of lesions. These measurements were stored and could later be extracted to draw a bounding box around each lesion. With these annotations a detection model or a weakly-supervised segmentation model can be trained.

\subsection{Universal Lesion Segmentation}
\label{sec:rw_uls}
Universal Lesion Segmentation (ULS) involves segmenting the various types of lesions that can be present within a specific anatomical region. These methods have historically focused on lesions in chest-abdomen-pelvis CT. ULS approaches may consist of a single end-to-end segmentation model, or they may include a classification step that first identifies the lesion type, and then selects a specialized model or adjusts pre- or postprocessing parameters automatically. ULS models can be used to automate the measurement of lesion diameters, reducing the inter-observer measurement variability of radiologists~\citep{moltz2012workflow}. Additionally, by incorporating a registration step, lesions can be tracked from baseline to follow-up scans~\citep{Cai2020_LesionTracker,hering2021whole} which can enhance the efficiency of radiological workflow~\citep{hering2024improving}. \\

ULS models need to generalize to infrequent lesion types for which little data is available in the public domain such as e.g. ovarian cancers or splenic lesions. They are often trained using weak supervision with pseudo-masks~\citep{zhou2023recist} due to the difficulty of collecting sufficiently large and diverse lesion datasets. \cite{Cai2018_SlicePropogated} first used pseudo-masks derived from diameter measurements to train a 2D ULS model with the DeepLesion dataset. They utilized the GrabCut~\citep{rother2004grabcut} algorithm for initial mask generation from the bidirectional measurements and showed enhanced performance through iterative training and prediction on slices lacking measurements. They reported a notable performance decrease when comparing single-slice to lesion volume segmentation, with Dice scores of 90.6\% and 76.4\%. \\

\cite{Tang2020_OneClick} introduced a single-click ULS network in 2020, eliminating the need for manual ROI delineation. \cite{agarwal2020weakly} investigated the use of co-segmentation in their ULS model. In subsequent studies Tang et al. studied incorporating regional level set loss~\citep{Tang2021_WeakSupervised}, attention mechanisms~\citep{Tang2021_Attention}, and finally developed \textit{MeaFormer} using a transformer architecture~\citep{tang2022accurate}. This model achieved the highest reported Dice scores for 2D and 3D segmentation, with 92.7\% and 85.6\% respectively, on the DeepLesion test set previously utilized by Cai et al. This test set consisted of 1000 manually 2D annotated lesions and a subset of 200 lesions which were fully annotated in 3D. Unfortunately, this test set is not available for benchmarking purposes, and the distribution of lesion sizes or types of lesions included remains undisclosed. \\

The ULS23 challenge aims to build upon the previous work in the field by granting researchers access to a diverse training dataset, along with a larger test set consisting of 725 3D annotated lesions spanning the chest-abdomen-pelvis area. Its goal is to enhance the performance of these models to a level where they can be used to reduce workloads in clinical practice.

\section{Materials}
\label{sec:mm}

\subsection{The ULS23 Dataset}
\label{sec:ds}

The training dataset for the challenge contains both fully 3D-segmented and partially 2D-annotated lesions. Whether to incorporate the partially annotated data into a semi-supervised learning method is at the discretion of the challenge participants. It should be noted that the lesion categories represented in the fully-annotated data do not cover the full range of lesion types which can be present in chest-abdomen-pelvis CT examinations. Models developed without utilizing the more diverse, partially-annotated data may exhibit limited generalizability to the test dataset. \\

Imaging data is provided to participants as volumes-of-interest (VOI's) of 256x, 256y, 128z voxels in the original scan spacing. Lesions with an axial diameter larger than this VOI were excluded from the test set. Where necessary, scans were padded with the minimum intensity value of the VOI minus one, allowing participants to establish where padding was added. The VOI's were sampled such that there always is a lesion voxel in the middle of the volume, simulating a click by a radiologist on that lesion. This voxel in the center of the volume was selected randomly from within the lesion mask. Within each VOI, there is only one annotated lesion. Any masks representing nearby lesions not connected to the central lesion were removed. The image and label data, provided in the NIfTI format, were preprocessed using the MONAI~\citep{cardoso2022monai} Python library. \\

For reproducibility and transparency, all code used to prepare the challenge data is available at the challenge GitHub repository~\citep{oncologyULS23}. For the 2D annotated data we also provide the 3D predictions of our best performing model, which can be used for further development of AI models. The ULS23 dataset is licensed under a Creative Commons Attribution-NonCommercial-ShareAlike 4.0 International License (CC BY-NC-SA 4.0).

\subsubsection{Training - Fully-Annotated Datasets}
\label{sec:training_fa}

\begin{figure}
  \centering
  \includegraphics[width=\textwidth]{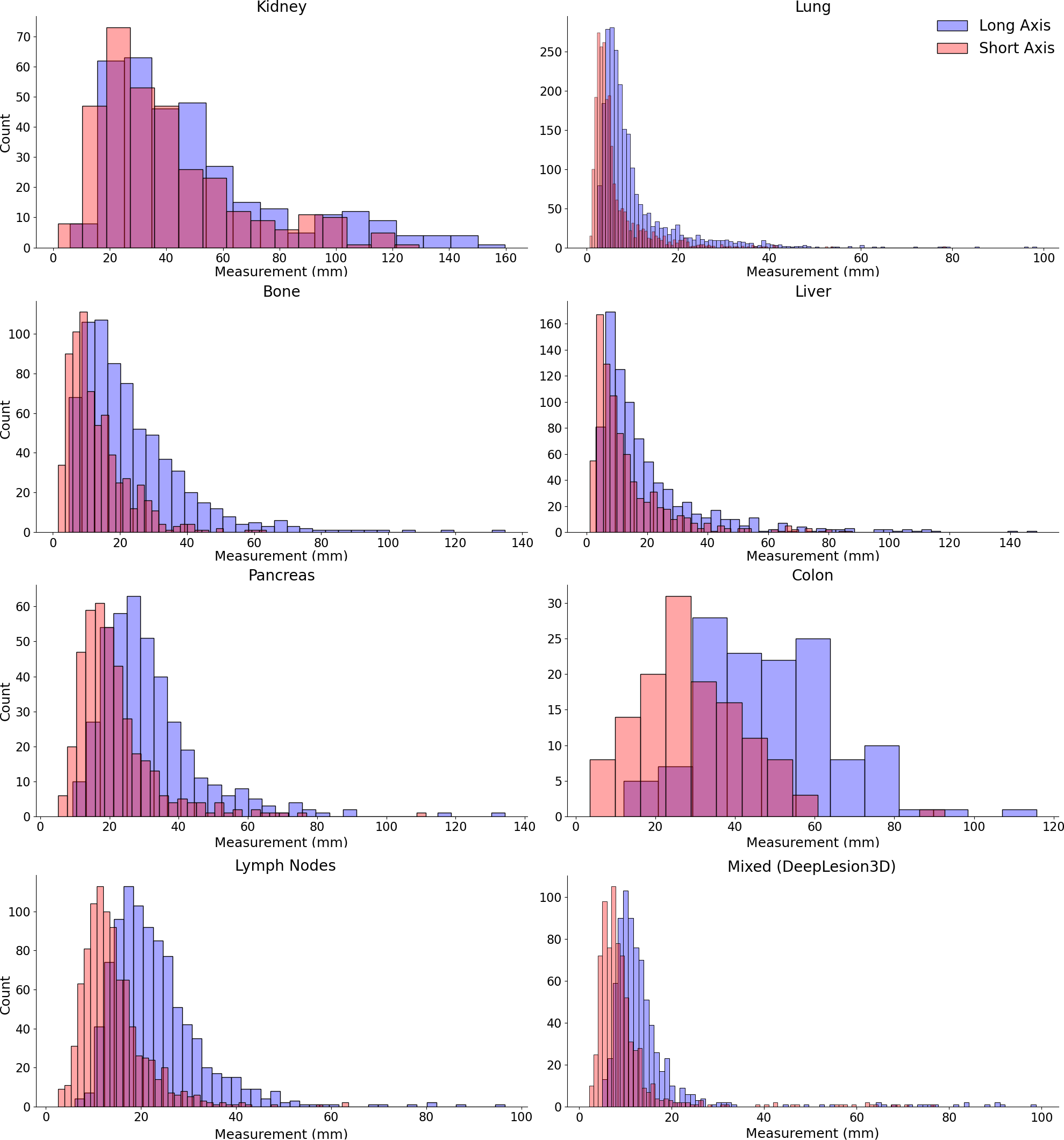}
  \caption{Histograms depicting the long- and short-axis measurements in millimeters for various lesion types in the fully-annotated training data reveal notable trends. Kidney and colon lesions tend to be larger on average. Lymph nodes, pancreas, and colon lesions exhibit a greater disparity between their long- and short-axis sizes, indicating that these lesions are more often non-spherical.}
  \label{fig:supervised_lesion_sizes_mm}
\end{figure}

The novel training data released for the challenge consists of 3D segmented lesions for a subset of the DeepLesion dataset, and two datasets containing bone and pancreas lesions~\citep{deGrauw2023uls23dataset}. Bone lesion have different morphological and intensity characteristics than soft-tissue lesions. As far as we are aware, there is no previous publicly available dataset with 3D segmentation masks for these kinds of lesions. The additional pancreatic lesion data is released with the aim of improving the segmentation performance for this notoriously difficult to segment lesion type~\citep{ghorpade2023automatic}. For the fully annotated data we also compiled eight public medical segmentation datasets containing a lesion segmentation component. See table \ref{tab:challenge_data} for an overview of the challenge training data. Figure \ref{fig:supervised_lesion_sizes_mm} contains the lesion size distribution per lesion type in the fully-annotated data in millimeters. Figure \ref{fig:novel_data_metadata} shows the scanner manufacturer, age and sex distributions for the novel data. Figure \ref{fig:additional_metadata} contains the study date and scan spacing distributions.

\begin{table}[H]
\centering
\resizebox{\textwidth}{!}{
\begin{tabular}{llrrrrrr}
\textit{Compiled Fully-Annotated Data} &  &  &  &  &  &  &  \\ \cline{1-6}
\textbf{Datasets} & \textbf{Data Type} & \textbf{Series} & \textbf{Lesions} & \textbf{Long-axis (pixels)} & \textbf{Short-axis (pixels)} & \textbf{} & \textbf{} \\ \cline{1-6}
KiTS21~\citep{heller2023kits21} & Kidney& 300 & 332 & $63.55 \pm 39.99$ & $51.30 \pm 32.39$ &  &  \\
LiTS~\citep{bilic2019liver} & Liver & 113 & 832 & $26.14 \pm 24.41$ & $17.17 \pm 16.62$ &  &  \\
NIH-LN ABD~\citep{roth2014new} & Lymph nodes & 85 & 557 & $25.79 \pm 11.78$ & $15.98 \pm \;\,8.08$ &  &  \\
NIH-LN MED~\citep{roth2014new} & lymph nodes & 90 & 375 & $32.08 \pm 12.75$ & $19.57 \pm \;\,8.52$ &  &  \\
LIDC-IDRI~\citep{armato2011lung,jacobs2016computer} & Lung & 750 & 2236 & $13.77 \pm 10.25$ & $8.23 \pm \;\,6.59$ &  &  \\
LNDb*~\citep{pedrosa2019lndb} & Lung & 208 & 692 & $11.95 \pm \;\,8.57$ & $7.05 \pm \;\,5.19$ &  &  \\
MDSC-Lung~\citep{antonelli2022medical} & Lung & 63 & 70 & $45.40 \pm 27.46$ & $27.21 \pm 16.87$ &  &  \\
MDSC-Colon~\citep{antonelli2022medical} & Colon & 126 & 131 & $63.20 \pm 23.78$ & $37.82 \pm 18.70$ &  &  \\
MDSC-Pancreas~\citep{antonelli2022medical} & Pancreas & 281 & 283 & $34.65 \pm 15.63$ & $24.31 \pm 12.72$ &  &  \\ \cline{1-6}
 &  &  &  &  &  &  &  \\
\textit{Novel Fully-Annotated Data} &  &  &  &  &  &  &  \\ \cline{1-6}
\textbf{Datasets} & \textbf{Data Type} & \textbf{Series} & \textbf{Lesions} & \textbf{Long-axis (pixels)} & \textbf{Short-axis (pixels)} & \textbf{} & \textbf{} \\ \cline{1-6}
Radboudumc-Bone~\citep{deGrauw2023uls23dataset} & Bone lesions & 151 & 697 & $31.17 \pm 21.50$ & $17.28 \pm 11.51$ &  &  \\
Radboudumc-Pancreas~\citep{deGrauw2023uls23dataset} & Pancreas & 119 & 120 & $56.14 \pm 23.98$ & $37.12 \pm 18.80$ & \textbf{} & \textbf{} \\
DeepLesion3D~\citep{Yan2018_DeepLesion} & Various lesions & 743 & 743 & $17.22 \pm 13.91$ & $11.69 \pm 10.48$ &  &  \\ \cline{1-6}
 &  &  &  &  &  &  &  \\
\textit{Partially-Annotated Data} &  &  &  &  &  &  &  \\ \cline{1-6}
\textbf{Datasets} & \textbf{Data Type} & \textbf{Series} & \textbf{Lesions} & \textbf{Long-axis (pixels)} & \textbf{Short-axis (pixels)} &  &  \\ \cline{1-6}
DeepLesion~\citep{Yan2018_DeepLesion} & Various lesions & 14498 & 31849 & $35.23 \pm 28.85$ & $25.25 \pm 22.87$ &  &  \\
CCC18~\citep{Urban2019CrowdsCureCancer} & Various lesions & 404 & 1211 & $51.81 \pm 43.29$ & $38.16 \pm 34.01$ &  &  \\ \cline{1-6}
 &  &  &  &  &  &  &  \\
\textit{Evaluation Data} &  &  &  &  &  &  &  \\ \cline{1-4}
\textbf{Datasets} & \textbf{Data Type} & \textbf{Series} & \textbf{Lesions} & \textbf{} & \textbf{} &  &  \\ \cline{1-4}
Validation set & Various lesions & 16 & 50 &  &  &  &  \\
Test set & Various lesions & 268 & 725 &  &  &  &  \\ \cline{1-4}
\end{tabular}
}
\caption{Overview of the data used in the ULS23 challenge. * The LNDb data licence does not allow repackaging their data, so it is not released as part of the training archive~\citep{deGrauw2023uls23dataset}. Instead, we release the code for participants to prepare the lesion VOI's for this dataset themselves~\citep{oncologyULS23}.}
\label{tab:challenge_data}
\end{table}

\subsubsection{Training - Partially-Annotated Datasets}
\label{sec:training_pa}

For semi-supervised training, we provide 2D pseudo-masks generated from the DeepLesion and Crowds Cure Cancer 2018 (CCC18)~\citep{Urban2019CrowdsCureCancer} datasets. The latter stems from a public annotation effort at the 2018 edition of the Radiological Society of North American (RSNA) meeting. For this dataset only those images with measurements made by a radiologist were used. Where multiple radiologists measured a single lesion, we randomly selected a measurement to create the pseudo-mask. \\

\begin{figure}[h]
  \centering
  \includegraphics[width=\textwidth]{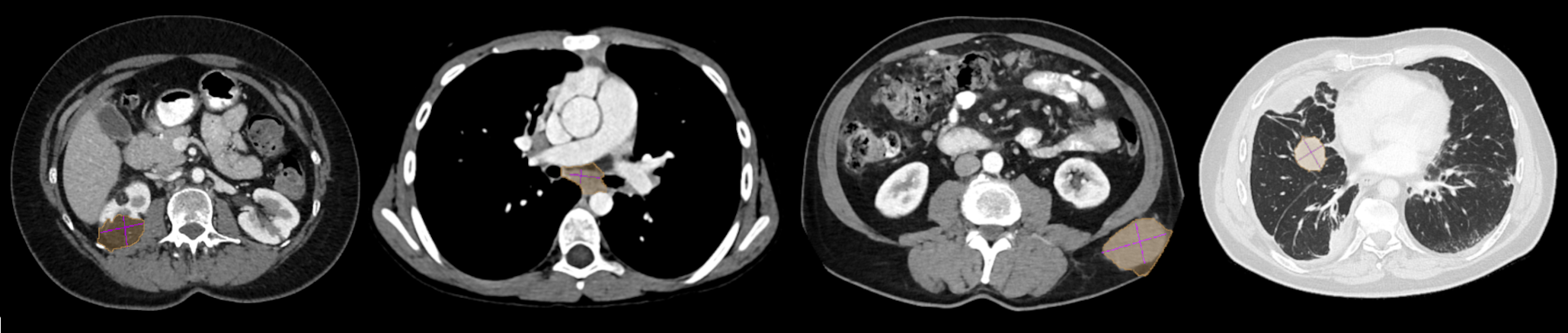}
  \caption{Examples of GrabCut pseudo-masks. From left to right, a kidney lesion, mediastinal lymph node, subcutaneous mass, and lung lesion. Note how GrabCut tends to oversegment (orange mask \textcolor{orange}{$\blacksquare$}) into healthy tissues compared to the reference measurements (purple lines \textcolor{custompurple}{$\blacksquare$}). Lung lesions are visualized using Window Level: -500 HU, Window Width: 1400 HU. Lesions outside the lungs with WL: 350 WW: 40.}
  \label{fig:grabcut_examples}
\end{figure}

Pseudo-masks were generated using the Grabcut algortihm~\citep{rother2004grabcut}, expanding on the approach of \cite{Cai2018_SlicePropogated}. GrabCut requires declaring four different image seeds: probable foreground (PFG), probable background (PBG), foreground (FG) and background (BG) regions. We used a dilation of the bounding box fitted to the RECIST measurements as PB. The area outside of this bounding box was set as BG. For the foreground definition we fit an ellipse to the lesion measurements. A dilation of this ellipse is used as PFG, whereas an erosion is set as FG. For each lesion we apply GrabCut multiple times. Initially, with the image normalized to the window level provided in the metadata. Subsequently, normalized to the average of the fitted elipse $\pm$ 50 and 100 Hounsfield units. We then chose from the GrabCut masks or directly utilized the ellipse fitted to the measurements, based on the calculation of the measurement error of each mask in comparison to the original measurement. Figure \ref{fig:grabcut_examples} shows examples of pseudo-masks extracted using GrabCut. As can be seen in these examples, using GrabCut frequently leads to over- and under-segmentation on challenging lesions.

\subsubsection{Test Dataset}
\label{sec:test}

The test set data was obtained from the PACS systems of the Radboudumc and Jeroen Bosch Ziekenhuis (JBZ) in the Netherlands. Natural Language Processing tools were used to analyze radiological reports and identify patients who had undergone imaging exams where target lesions were measured. Each radiological report contained a short description of the lesion, the axial slice and series identifier it was located on, and the size measured by the radiologist during clinical practice. To identify and segment the lesions, six annotators with a (bio)medical background were instructed to locate and remeasure the lesion indicated in the selected reports. Subsequently, an experienced radiologist (with over 10 years of experience) reviewed and, where necessary, corrected the measurements. The annotators then segmented the lesions using these measurements as a guide. For each lesion, three annotators segmented it in 3D, and the resulting majority vote was reviewed, and corrected where necessary, by a radiologist. The test set contains a diverse range of lesion types, including but not limited to, enlarged lymph nodes, and lesions in the kidney, colon, pancreas, bones, lung, liver, peritoneum, and mammae. \\

The test set data comprises 284 patients, 152 from the Radboudumc and 132 from the JBZ. Of these patients, 53.5\% are male and 46.5\% are female. In total, there are 775 lesions in the test set, 368 from Radboudumc and 407 from JBZ. The radiological report explicitly mentioned a lesion being measured as a baseline for 51.4\% of lesions. After separation of the validation set used during the development phase of the challenge, the final test set consists of 268 patients and 725 lesions. The age distribution per sex, and scanner manufacturer distribution is shown in figure \ref{fig:novel_data_metadata}.

\subsubsection{Validation Dataset}
\label{sec:validation}
A portion of the test data was allocated to create a validation set, containing 50 lesions from 16 distinct patients, with an equal distribution of 25 lesions from each center in the test set. These lesions were carefully chosen from the test dataset to represent a diverse range of lesion types and sizes. It is important to note that there is no overlap of patients between the validation set and the test set, ensuring the integrity of the evaluation process.

\subsection{Challenge Design}
\label{sec:rules}
The ULS23 challenge is a Type II challenge, requiring containerized solutions which can automatically be run on novel data. This ensures the test set scans do not need to be distributed to participants and allows for automatic processing of submissions for benchmarking purposes after the conclusion of the primary event. During the final evaluation of the challenge phase, participants were permitted only a single successful submission, preventing fine-tuning on the test set. Instead, participants were encouraged to test their methods on the validation set to ensure that their containerized algorithms function properly. The models provided by the challenge organizers were developed and evaluated using only the training data. The baseline model was only evaluated once on the test set, during the final challenge evaluation.

\subsubsection{Participation Requirements}
Each participant was required to join or create a team under which they participated in the challenge. Teams may consist of an arbitrary number of participants. The challenge participants agreed that their algorithms will be made publicly available for research purposes on the grand-challenge.org platform~\citep{Meakin_Grand-Challenge_org}. To be eligible to win the challenge, participants were required to publicly release a short manuscript describing their methodology, including architecture type, model size, loss functions and hyperparameters used. Participants were allowed to use external data when developing their models, including data not publicly available, as long as they provide a summary description of the amount and types of new lesions, and the number of patients and series in the manuscript detailing their methodology. These documents were archived by the challenge organizers for future reference by other researchers. Members of the organizers' institutes may also submit their solutions to the challenge but are not eligible to win the challenge.\\

After the primary event has concluded, submitting to the leaderboard no longer requires open-sourcing algorithms, making them publicly accessible on Grand Challenge or disclosing the methodology, although we do strongly encourage this. By removing these constraints institutions or participants with commercial interest can also evaluate their models on the test set, and this could provide important insights into the performance of commercial versus research algorithms. 

\subsubsection{Timeline \& Results}
On October 29th, 2023 the training data was released, marking the start of the challenge. The validation leaderboard was opened on February 1st, 2024. Submission to the challenge test set was allowed starting from March 17th, 2024. On April 9th, 2024 the primary event of the challenge was concluded, at which point the top-3 teams with an algorithm beating the baseline model were announced, in their respective positions, as the winners of the ULS23 challenge. Up-to 3 authors of each of these winning teams were invited to be co-authors on the challenge evaluation paper. After this date the challenge remains open to submissions for benchmarking purposes.

\subsubsection{Infrastructure}
\label{sec:infrastructure}
We used the Grand Challenge platform to host the challenge documentation, leaderboard, and for submitting algorithms for inference on the test and validation set. The ULS23 website is accessible at \url{https://uls23.grand-challenge.org}. The platform also provides users across the world direct access to the algorithms developed during the challenge. The baseline model weights~\citep{deGrauw2023uls23model} and the challenge image data are stored as archives on Zenodo~\citep{deGrauw2023uls23dataset}. The code used to prepare the ULS23 training dataset, the evaluation software and run the baseline model in a containerized manner is hosted on GitHub~\citep{oncologyULS23}. The annotations for the training set and baseline model predictions are also hosted on GitHub, allowing anyone to create an issue if there are annotation errors, and ensuring that any changes made to the labels are archived. \\

To submit to the challenge, participants need to create a docker container containing a fully automatic inference pipeline for their algorithm, including pre- and postprocessing of images. It is not possible to submit semi-automated or interactive methods. We have released documentation on how to construct such a container on the challenge website. The maximum compute available to participants during inference is a NVIDIA T4 GPU with 16GB of VRAM, 8 vCPU and 32GB of CPU RAM. Each job has a maximum runtime of 9 minutes, during this time the whole inference pipeline from loading the model and images to exporting the output predictions must be run for 100 lesions. By restricting the inference time for each lesion volume, we mimic the time constraints posed on ULS algorithms when used in an online-setting by Radiologists reading scans during clinical practice. Furthermore, this encourages participants to deliver optimized solutions, by effectively using the available data and compute resources, while discouraging the use of overly large ensembles of multiple models.

\subsubsection{Evaluation}
\label{sec:evaluation}
In order to win the challenge, participants need to optimize for three components: the segmentation performance, inference speed and segmentation consistency. We opt for this combined evaluation scheme instead of focusing solely on segmentation performance in order to ensure that the submitted models remain applicable in a clinical setting, where inference time and compute is limited and robust models are essential. \\

For the average segmentation performance (SP), we evaluate the predicted 3D masks against the reference segmentation using the Sørensen-Dice coefficient. We also calculate the symmetric mean absolute percentage error for the long- and short-axis measurements (LAE and SAE). Finally, a subset of lesions is included multiple times during evaluation of the validation and test set, using randomly sampled lesion foreground voxels as the center locations. This results in slight variations on the scan context for each cropped VOI. We check whether the model outputs similar predictions using these different click locations by comparing the Sørensen–Dice coefficient of the re-aligned segmentation masks. A model robust to click location variation should have a high average segmentation consistency score (SCS). \\

The final challenge score (CS) used to rank submissions can be expressed as:
\begin{equation}
    \text{CS} = 0.8 \cdot \text{SP} + 0.05 \cdot \text{LAE} + 0.05 \cdot \text{SAE} + 0.1 \cdot \text{SCS}
\end{equation}

where LAE and SAE are calculated using the Symmetric Mean Absolute Percentage Error (SMAPE):
\begin{equation}
\text{SMAPE} = \frac{1}{n} \sum_{i=1}^{n}{\frac {|\hat{y}_{i}-y_{i}|}{|\hat{y}_{i}|+|y_{i}|}}
\end{equation} \\

The challenge score is used to determine the winners of the challenge. In the challenge evaluation paper we will test whether the differences in performance of the models developed during the challenge are significant, and will compare them against submissions from the open leaderboard phase. 

\section{Methods}

\label{sec:baselines}
In conjunction with the ULS23 challenge, we developed a baseline model using the challenge dataset and the LNDb data. To assist participants in preparing their algorithms for the challenge infrastructure, we released the model weights, training code and algorithm container. Additionally, the algorithm can be accessed on the Grand Challenge platform\footnote{\url{https://grand-challenge.org/algorithms/universal-lesion-segmentation-uls23-baseline/}}, where users can upload their own data to be segmented.

\subsection{Training Approach}
We utilized the nnU-net framework~\citep{isensee2021nnu} (version 2; January 9, 2024) for testing and developing our baseline model. We compared our modifications with the standard nnU-net configuration provided by the framework. Given the specific characteristics of our dataset, where a single lesion requiring segmentation is positioned at the center of each Volume of Interest (VOI), we found it necessary to make two adjustments to the default configuration. First, we disabled the standard resampling function of the nnU-net framework to preserve the original size of the VOI, regardless of spacing. Then, we increased the patch size to encompass the entire VOI, measuring 256x, 256y, 128z voxels. These adjustments ensured that no patching or resampling alters the implicitly encoded, spatial information regarding the lesion's central location within the VOI. \\

\begin{figure}[h]
  \centering
  \includegraphics[width=\textwidth]{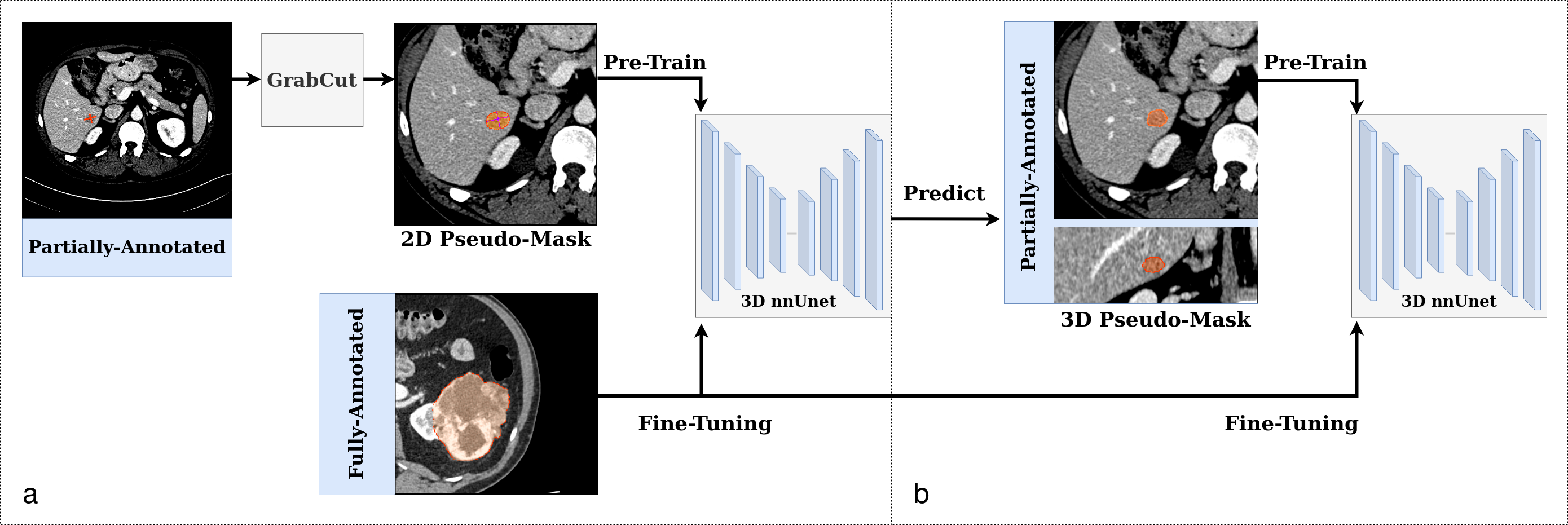}
  \caption{Training pipeline for the semi-supervised baseline model. a) In the first training iteration a nnUnet is pretrained using the 2D GrabCut masks generated from the partially annotated data, and then fine-tuned on the fully annotated data. b) In the second training iteration a different nnUnet is pretrained using the predicted 3D pseudo-masks for the partially annotated data and then fine-tuned using the fully-annotated data.}
  \label{fig:training_pipeline}
\end{figure}

The majority of the challenge data is partially annotated and not directly suited for training a 3D segmentation model. Using only the fully annotated data however, could lead to a model that does not generalize as well to other lesion types or data sources. Therefore, our training pipeline aimed to leverage the partially labeled data from the DeepLesion and CCC18 datasets through the use of pseudo-masks and iterative training, as visualized in figure \ref{fig:training_pipeline}. \\

For this semi-supervised model, we selected a 3D residual encoder architecture that includes an additional layer in both the encoder and decoder sections, along with enlarged feature dimensions. These modifications relative to the automatically configured plans were intended to fully use the computational capabilities of the NVIDIA A100 40GB GPU available during training. We first pre-trained the model for 1000 epochs on the 2D GrabCut masks from the partially annotated data. It was then fine-tuned for 500 epochs at 25\% of the original learning rate using the fully annotated data. Using this first model, we then predicted 3D pseudo-masks for the partially annotated data. We then selected from these predictions, those cases where the long- and short-axis error is 5 pixels or less, leaving us with 18,264 lesion to pre-train a second nnUnet. This model was subsequently fine-tuned using the fully-annotated data, resulting in the final  (nnUnet-ResEnc+SS). We compare this training pipeline against an identical residual encoder nnUnet trained in a single 1000 epoch run using only the fully-annotated data (nnUnet-ResEnc), and against an automatically configured nnUnet. Detailed model hyperparameters and data normalization properties are discussed in \ref{sec:cds:appendixB2}. \\

All three models were first evaluated on held-out training data consisting of 10\% of each fully annotated training dataset, split on a patient level. We grouped the results from these datasets to report the average performance per lesions type. A list of the cases used for this evaluation is available in \ref{sec:cds:appendixB3}. Before evaluating on the test set, we retrained the models using all of the available training data. 

\subsection{Results}
\label{sec:results}
Table \ref{tab:results} shows the mean Dice segmentation performance and the standard deviation across various configurations for both the held-out training data and the challenge test set. We report the results for each lesion type from the fully annotated data. Additionally, for the test set, we also include results for the lesion types that were not seen during training.

\begin{table}[H]
\resizebox{\textwidth}{!}{
\begin{tabular}{ccccccccc}
 &  &  &  &  & Dice &  &  &  \\ \hline
 & \multicolumn{1}{c|}{} & \textbf{Kidney ($N=33$)} & \textbf{Lung ($N=267$)} & \textbf{Lymph-Node ($N=77$)} & \textbf{Bone ($N=86$)} & \textbf{Liver ($N=54$)} & \textbf{Pancreas ($N=38$)} & \textbf{Colon ($N=12$)} \\ \hline
Held-Out & \multicolumn{1}{l|}{nnUnet} & $0.617 \pm 0.249$ & $0.753 \pm 0.137$ & \underline{$0.698 \pm 0.174$} & \underline{$0.674 \pm 0.262$} & $0.640 \pm 0.173$ & $0.556 \pm 0.221$ & $0.493 \pm 0.236$ \\
Training Data & \multicolumn{1}{l|}{nnUnet-ResEnc} & $0.718 \pm 0.244$ & $0.751 \pm 0.147$ & $0.698 \pm 0.181$ & $0.637 \pm 0.275$ & $0.642 \pm 0.173$ & $0.605 \pm 0.214$ & $0.394 \pm 0.250$ \\
 & \multicolumn{1}{l|}{nnUnet-ResEnc+SS} & \underline{$0.724 \pm 0.257$} & \underline{$0.754 \pm 0.149$} & $0.698 \pm 0.182$ & $0.648 \pm 0.267$ & \underline{$0.650 \pm 0.184$} & \underline{$0.616 \pm 0.196$} & \underline{$0.518 \pm 0.208$} \\
 &  &  &  &  &  &  &  &  \\ \hline
 & \multicolumn{1}{c|}{} & \textbf{Kidney}* & \textbf{Lung} & \textbf{Lymph-Node} & \textbf{Bone}* & \textbf{Liver} & \textbf{Pancreas}* & \textbf{Colon}* \\ \hline
 \multirow{2}{*}{Test Set} & \multicolumn{1}{l|}{nnUnet} & $0.574 \pm 0.240$ & $0.719 \pm 0.243$ & $0.646 \pm 0.230$ & $0.289 \pm 0.226$ & $0.620 \pm 0.258$ & $0.555 \pm 0.164$ & $0.437 \pm 0.205$ \\
 & \multicolumn{1}{l|}{nnUnet-ResEnc} & \underline{$0.691 \pm 0.195$} & $0.753 \pm 0.222$ & \underline{$0.685 \pm 0.220$} & $0.413 \pm 0.256 $ & $0.706 \pm 0.238$ & \underline{$0.756 \pm 0.094$} & $0.430 \pm 0.094$ \\
 & \multicolumn{1}{l|}{nnUnet-ResEnc+SS} & $0.633 \pm 0.197$ & \underline{$0.761 \pm 0.211$} & $0.685 \pm 0.233$ & \underline{$0.419 \pm 0.257$} & \underline{$0.730 \pm 0.214$} & $0.707 \pm 0.123$ & \underline{$0.541 \pm 0.211$} \\
 &  &  &  &  &  &  &  &  \\ \cline{1-5}
 & \multicolumn{1}{c|}{} & \textbf{Full Test Set ($N=725$)} & \textbf{FSUP ($N=630$)} & \textbf{PSUP ($N=95$)} &  &  &  &  \\ \cline{1-5}
 \multirow{2}{*}{Test Set} & \multicolumn{1}{l|}{nnUnet} & $0.651 \pm 0.253 $ & $0.659 \pm 0.246$ & $0.600 \pm 0.290$ &  &  &  &  \\
 & \multicolumn{1}{l|}{nnUnet-ResEnc} & $0.700 \pm 0.235$ & $0.708 \pm 0.228$ & \underline{$0.648 \pm 0.275$} &  &  &  &  \\
 & \multicolumn{1}{l|}{nnUnet-ResEnc+SS} & \underline{$0.703 \pm 0.240$} & \underline{$0.714 \pm 0.226$} & $0.631 \pm 0.305$ &  &  &  & 
\end{tabular}
}
\caption{Segmentation performance comparison on the 10\% held-out training data per lesion type and the test set. For the individual lesion types in the test set, * indicates there were $\leq$ 20 lesions of this type in the test set. The exact distribution of lesion types is not provided to participants. FSUP = fully supervised lesions types (i.e. Kidney - Colon). PSUP = lesion types present in the partially supervised training data.}
\label{tab:results}
\end{table}

\begin{figure}[h]
  \centering
  \includegraphics[width=\textwidth]{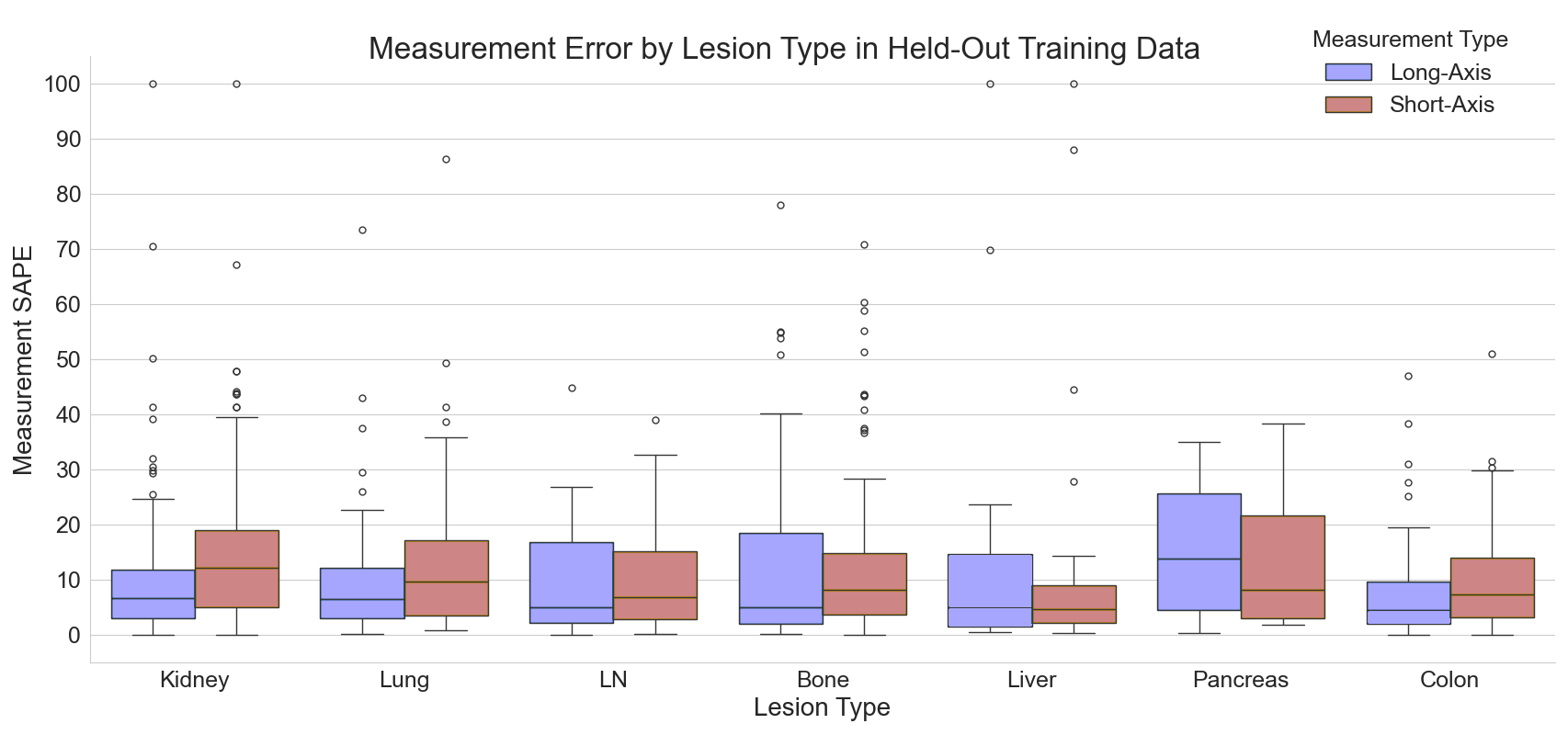}
  \caption{Boxplots of the long- and short-axis measurement errors for the baseline model on the different lesion types in the held-out training data. SAPE = Symmetric Average Prediction Error.}
  \label{fig:measurement_error_distributions}
\end{figure}

\begin{figure}[h]
  \centering
  \includegraphics[width=0.75\textwidth]{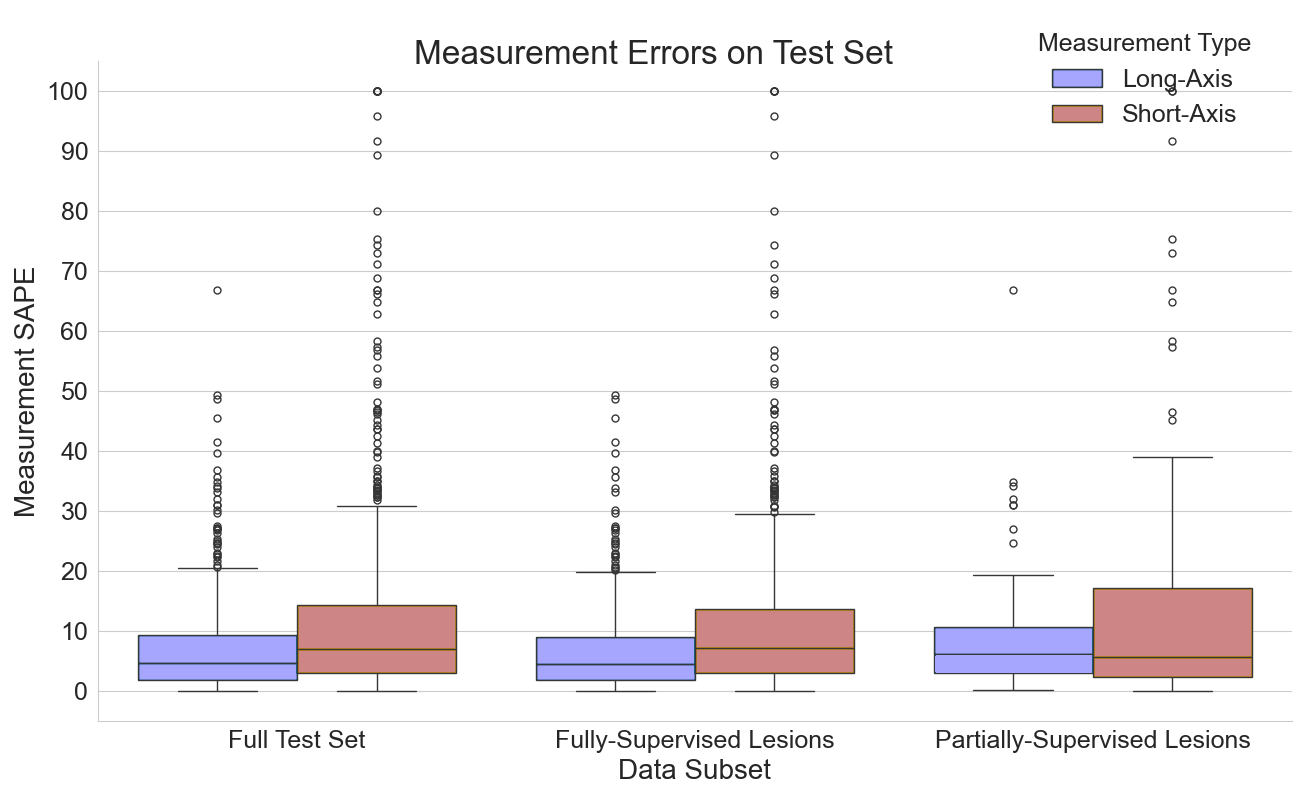}
  \caption{Boxplots of the long- and short-axis measurement errors for the baseline model on the test set. The fully-supervised types are lung, liver, kidney, colon, pancreas, bone lesions and lymph nodes. Partially-supervised lesion types are those included in the partially annotated data e.g. adrenal, ovary, subcutaneous. SAPE = Symmetric Absolute Percentage Error. }
  \label{fig:measurement_error_distributions_test}
\end{figure}

Figure \ref{fig:measurement_error_distributions} shows the long- and short-axis axial measurement error distributions for the different lesion types in the held-out training data as predicted by the semi-supervised model. On the test set it achieves an overall ChallengeScore of 0.729, consisting of a mean Dice score of $0.703 \pm 0.240$, long-axis SMAPE of $11.2\% \pm 15.8\%$, short-axis SMAPE of $12.0\% \pm 15.9\%$, and consistency Dice score of $0.787 \pm 0.252$. Figure \ref{fig:measurement_error_distributions_test} shows the measurement error distribution for the full test set, and for those lesions types which were or were not contained in the fully-annotated training data. 

\section{Discussion}
The ULS23 challenge represents a major step forward in the assessment of universal lesion segmentation models for computed tomography, featuring the most extensive and diverse set of 3D annotated lesions to date. By selecting clinically relevant 'target' lesions for our test set, and prioritizing fast, lightweight models, we have aligned our efforts closely with real-world clinical needs. This allows the challenge to serve as a strong benchmark for current and future universal lesion segmentation models. \\

Our evaluation of the three model configurations trained on the ULS23 challenge data shows robust performance in terms of segmentation and measurement accuracy. Notably, our best model achieved a long-axis measurement error within 10\% of the reference lesion size in 69.2\% of cases. For the short-axis measurements, this number stands at 64.8\%. In comparison, a previous meta-analysis~\citep{yoon2016observer} on the intra-observer variability of 1D measurements of RECIST lesions reported a pooled mean relative measurement difference (RMD) of 1.6\% (with a 95\% confidence interval (CI) of -2.8 to 6.0) for the long-axis of lesions. The pooled standard deviation (SD) of this RMD was 12.1\% (with a 95\% CI of 8.6 to 15.6). This indicates that there is a similar degree of measurement variability between radiologists. Further research is required to determine the acceptable levels of segmentation and measurement variability that would enable these models to accelerate clinical processes. \\

\begin{figure}
  \centering
  \includegraphics[width=\textwidth]{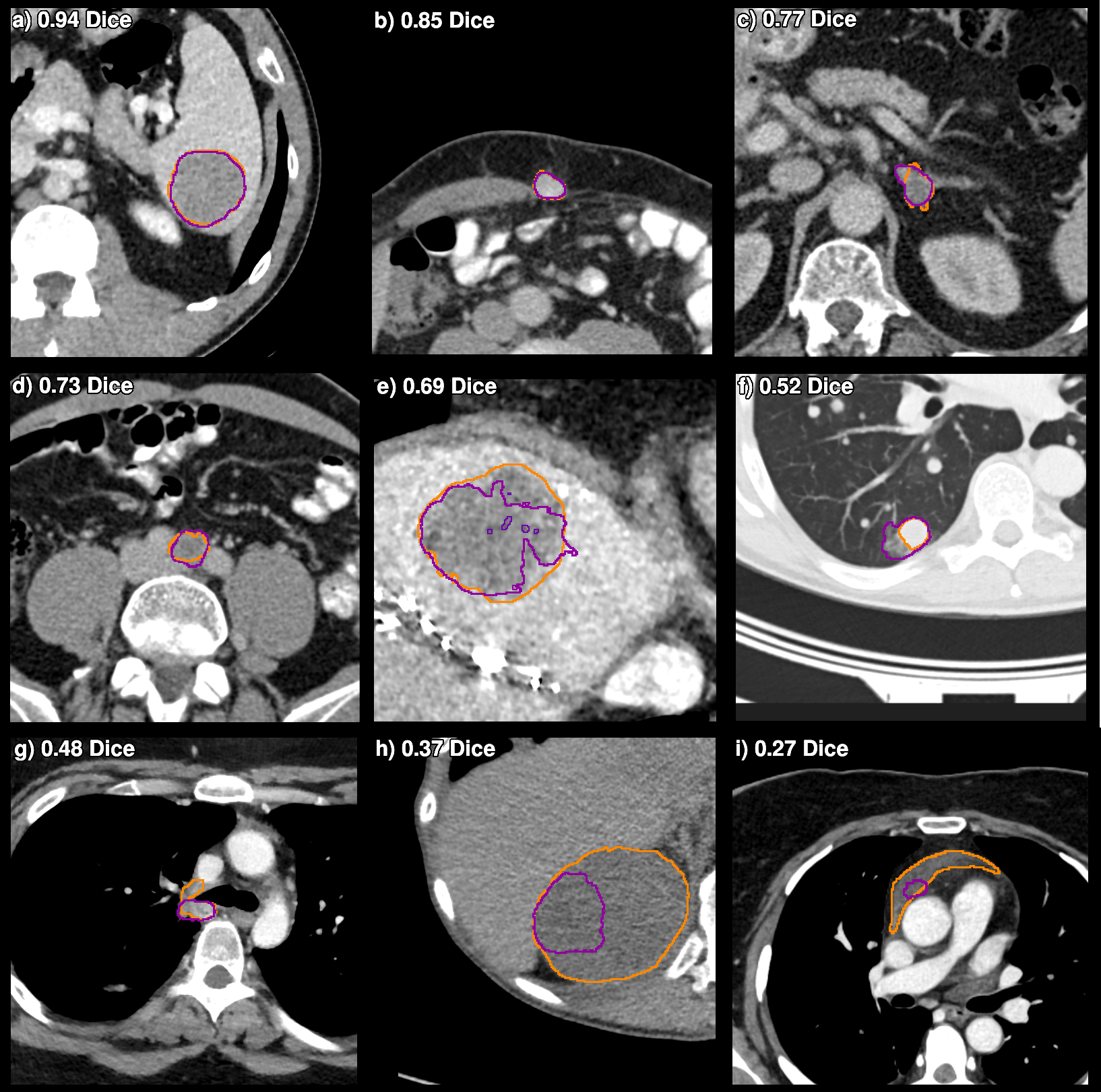}
  \caption{Ground truth (orange line \textcolor{orange}{$\blacksquare$}) and baseline model prediction (purple line \textcolor{custompurple}{$\blacksquare$}) on axial slices from the test set. The 3D Dice score for each lesion is included in the top-left corner. The lesions visualized are: a) spleen lesion b) lesion in the abdominal wall c) adrenal lesion d) abdominal lymph node e) liver lesion f) lung lesion g) mediastinal lymph node h) kidney lesion i) Pericardial lesion. Lung lesions are visualized using Window Level: -500 HU, Window Width: 1400 HU. Lesions outside the lungs with WL: 350 WW: 40.}
  \label{fig:test_set_result_images}
\end{figure}

The results presented in Table \ref{tab:results}, demonstrate considerable variations in segmentation performance across different lesion types, a finding consistent with outcomes from previous segmentation challenges~\citep{antonelli2022medical, bilic2019liver, heller2023kits21, pedrosa2019lndb}. Lesions in the colon or pancreas are particularly challenging to segment due to their low contrast with surrounding healthy tissue and their irregular shapes. Figure \ref{fig:test_set_result_images} illustrates the variation in model performance across nine test cases. Lesion $h$ and $i$ highlight the issue of under-segmentation of larger lesions, which is a consistent weakness of the evaluated models. One possible explanation is that larger lesions appear less frequently in the training data (see Figure \ref{fig:supervised_lesion_sizes_mm}). Future work could try to address this problem by over-sampling large lesions during training. Lesion $f$ is an example of the inherent subjectivity of certain lesion boundaries, with the radiologist only segmenting the solid component of the lung lesion and the algorithm including the ground-glass area. Similar subjectivity can also be observed in merging liver lesions or lymph nodes. Future iterations of the challenge should explore how to minimize this subjectivity. \\

We found that using a residual encoder architecture, scaled to the available compute resources, outperformed the standard nnUnet configuration on the test set. This finding has been corroborated by recent work from Isensee et al.~\citep{isensee2024nnunet}. For the held-out training data, we observed a slight improvement in segmentation performance for lesion types with limited fully-annotated examples, such as kidney, pancreas, and colon lesions, when including weakly-annotated lesions during training. Interestingly, on the test set, the residual encoder nnUnet trained solely on the eight fully annotated lesion types showed a slight performance advantage for lesion types that were not included in the fully annotated data (PSUP). However, further analysis revealed that the semi-supervised model also performed better for this category when using test time augmentation, as discussed in \ref{sec:cds:appendixB1}. Given that the semi-supervised model had the best overall performance, we chose it as the baseline for the challenge. \\

In Figures \ref{fig:error_plots_held_out} and \ref{fig:error_plots_test}, we show how the measurement error of the long- and short-axis of a lesion correlates with the Dice score of each prediction. While higher Dice scores are generally associated with lower measurement errors, there are also exceptions where deviations from the reference segmentation happen to result in measurements that are close to the actual lesion size. Furthermore, we observed that smaller lesions, such as those in the lungs or bones, can show significant relative measurement errors despite a high Dice score due to minor differences in the shape of the predicted masks. These findings highlight the challenge in generalising the results from previous studies~\citep{Cai2018_SlicePropogated,Tang2020_OneClick} which report measurement errors based on 2D segmentation, and comparing them to 3D segmentation performance. In contrast to previous studies~\citep{Tang2021_Attention,tang2022accurate}, we chose not to assess the measurement error in millimeters. Direct evaluation of the measurement error can underestimate the true error when smaller lesions predominate in a subset of the data, complicating comparisons across different lesion types. Furthermore, the measurement error in millimeters is influenced by the resolution of the scans used in the evaluation. \\

Previous studies used DeepLesion annotations to assess the 2D segmentation performance and measurement error specifically on the axial slice where the lesion reaches its maximum extent. On these slices, lesions are often easier to differentiate from surrounding tissues due to their larger size and the reduced partial volume effect~\citep{Souza2005}, compared to slices where lesions start or end. The task addressed by the ULS23 challenge is inherently more complex, since any slice could be sampled as the center of the volume of interest through a simulated click annotation. The advantage of this approach lies in the fact that it eliminates the need for identifying the correct slice when manually selecting a lesion. During evaluation, we observed that changing the center coordinate of the lesion's volume of interest affects the model's prediction, as indicated by the consistency Dice score. This ``intra-observer variability" of the model should be considered when applying or evaluating a click- or detection-guided ULS model in a clinical setting. \\

The ULS23 challenge is constrained by a number of limitations. For the training data, we compiled datasets from multiple sources, resulting in a heterogeneous collection which introduces extensive variability in patient demographics, scanner manufacturers and image acquisition protocols. This heterogeneity typically enables models to better handle out-of-distribution data. However, it can also introduce significant variations in annotation quality. Annotations in the training data range from expert-level precision to images with imprecise lesion boundaries. Furthermore, the collected datasets primarily contain semantic annotations, which may not be ideally suited for tasks requiring precise instance segmentation such as for this challenge. This incompatibility may lead to suboptimal model performance when handling adjacent lesions. The LiTS dataset is noteworthy for containing many merging and adjacent lesions where it is hard to establish individual lesion boundaries. Considerable performance increases might be gained by manually or heuristically splitting these lesions. \\

Another limitation of the challenge data is that the lesions in the test set were annotated by a single radiologist, which can lead to a narrow view of segmentation performance by not accounting for inter-observer variability. Due to the significant annotation burden associated with collecting 3D annotations, we chose to prioritize the number of lesions annotated for this iteration of the challenge. As a result, our test set is over three times larger than the previously used set of 3D-annotated DeepLesion cases, which was also collected using a single reader~\citep{Cai2018_SlicePropogated}. Nevertheless, future iterations of the challenge could benefit from collecting additional annotations per lesion to study the inter-observer variability of radiologists.\\

The ULS23 challenge was run in a Type II format, allowing the test set to remain fully hidden, making it a reliable long-term benchmark. However, this requires participants to containerize their solutions to ensure compatibility with the challenge's infrastructure. This presented a higher entry barrier to participants. To mitigate this, we provided documentation and hosted an information session on preparing algorithms for the challenge platform. Participants of the challenge were also required to submit a short paper describing their methodology and any external data used. Although we did not mandate that participants open-source their solutions, we strongly encouraged this. Instead, participants needed to make their algorithms publicly available through Grand Challenge for use on novel data. By adopting this approach, we prioritized providing researchers access to high-quality ULS algorithms, sacrificing some transparency in model development in the process. \\

Finally, the computational resources available through the Grand Challenge platform prevented the evaluation of more demanding, and potentially more accurate models. Partially due to this limitation, we decided to focus this first iteration of the challenge on smaller models with faster inference times. Leading to models that are intended for use in online settings by radiologists, offering a practical balance between performance and compute requirements.

\section{Conclusions}
\label{sec:concl}
This paper presents the ULS23 challenge, establishing the first public benchmark for the evaluation of 3D universal lesion segmentation models on computed tomography scans. We introduce novel training data for bone and pancreas lesions for which only limited public data was previously available. The challenge training dataset features a unique combination of fully- and partially-annotated data. To demonstrate the potential of this combined dataset, we developed a strategy for predicting 3D pseudo-masks from the partially-annotated 2D data, allowing for their inclusion in 3D model development. Using this approach, we iteratively trained a semi-supervised ULS model that leverages the entire training dataset. For evaluation purposes, we assembled a high-quality and diverse test set of lesions that were selected for RECIST measurement in clinical practice. By focusing on clinically relevant target lesions, our benchmark is tightly integrated with the practical requirements of radiologists. Our scaled-up, semi-supervised model achieves a Dice score of $0.703 \pm 0.240$ on this test set, compared to a Dice score of $0.651 \pm 0.253$ for a standard, automatically-configured nnUnet. The model weights, data processing code and evaluation scripts are publicly released to ensure transparency and reproducibility. Future work will include a meta-analysis of the methods developed by challenge participants and an assessment of how these models could reduce reading times for oncological scans. Subsequent iterations of the challenge can explore various aspects of ULS model development such as prioritizing segmentation performance over inference speed, expanding evaluation on rare lesion types or including different imaging modalities.

\section{Declaration of Competing Interest}
\label{sec:interests}
B.v.G. is founder and shareholder of Thirona. M.P. receives grants from Canon Medical Systems, Siemens Healthineers; royalties from Mevis Medical Solutions; and payment for lectures from Canon Medical Systems and Siemens Healthineers. The host institution of MP is a minority shareholder in Thirona. B.v.G. and M.P. report no other relationships that are related to the subject matter of the article. The other authors declare that they have no known competing financial interests or personal relationships that could have appeared to influence the work reported in this paper.

\section{Acknowledgments}
\label{sec:acknowledgments}
Compute costs for the ULS23 Challenge were sponsored by Amazon Web Services via the Radboudumc. We are grateful to Grand Challenge for providing an excellent platform to host our challenge, and the algorithms developed during it. Additionally, we would like to thank R. Summers and T. Mathai for their advice and support regarding the DeepLesion dataset, and the numerous contributors who meticulously curated the various datasets included in the challenge training data.

\section{CRediT Authorship Contribution Statement}
\textbf{M.J.J. de Grauw}: Conceptualization, Data curation, Methodology, Software, Investigation, Formal analysis, Writing – original draft, Visualization. \textbf{E.Th. Scholten}: Data curation, Writing - Review \& Editing. \textbf{E.J. Smit}: Data curation, Writing - Review \& Editing. \textbf{M.J.C.M. Rutten}: Data curation, Writing - Review \& Editing. \textbf{M. Prokop}: Conceptualization, Writing - Review \& Editing. \textbf{B. van Ginneken}: Conceptualization, Writing - Review \& Editing, Project administration. \textbf{A. Hering}: Conceptualization, Writing - Review \& Editing, Supervision.

\section{Declaration of Generative AI in Scientific Writing}
During the preparation of this work the authors used generative AI in order to improve readability and language. After using this tool/service, the authors reviewed and edited the content as needed, and take full responsibility for the content of the publication. 

\bibliographystyle{abbrvnat}
\bibliography{library_cleaned_short_nourl}

\appendix

\section{Supplementary Materials}
\label{sec:cds:appendixB}

\begin{figure}[h]
  \centering
  \includegraphics[width=\textwidth]{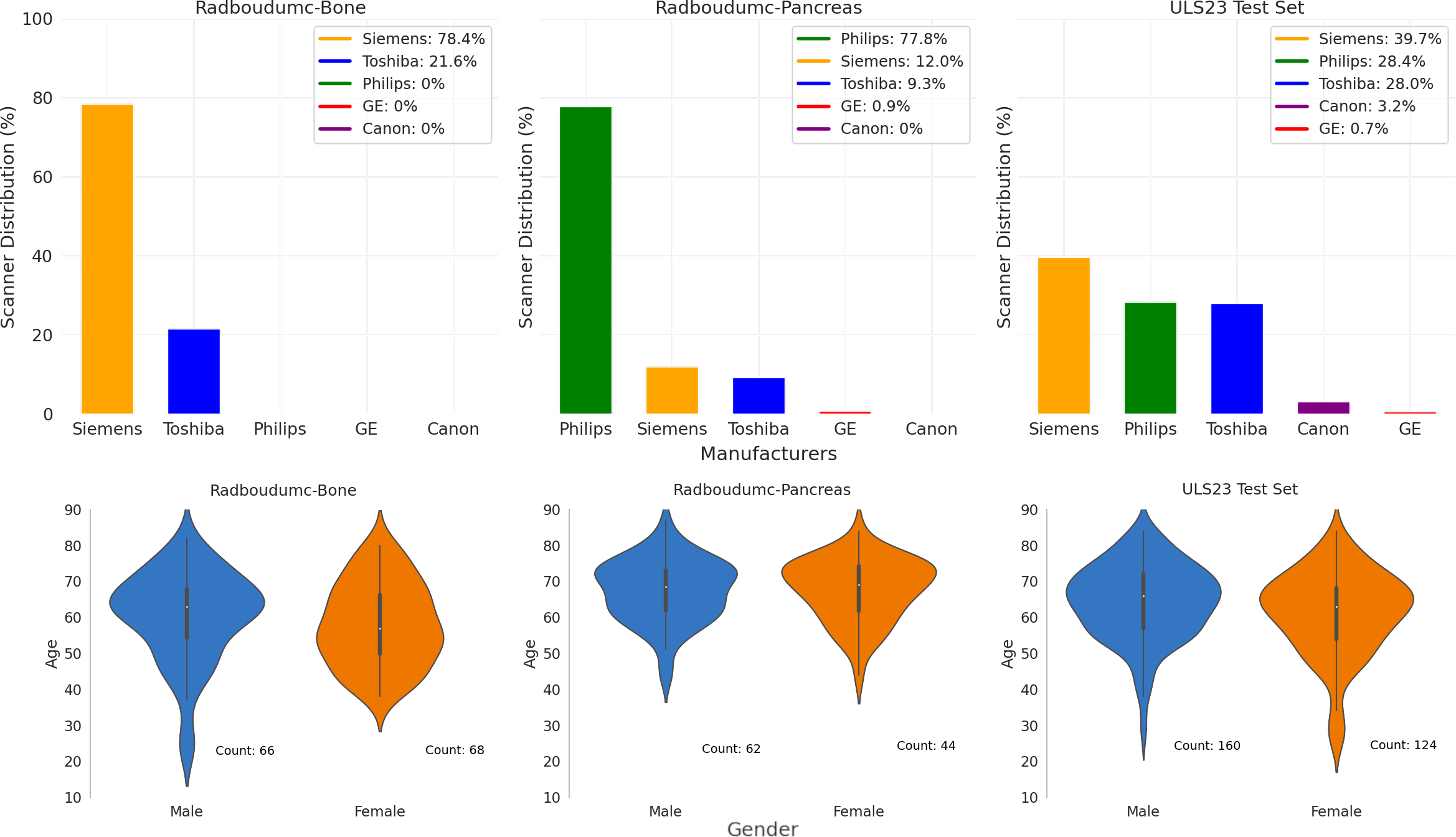}
  \caption{Age, sex and scanner manufacturer characteristics of the novel training data and the test set. For 3 series of the Radboudumc-Bone dataset and 13 series of the Radboudumc-Pancreas dataset the metadata could not be recovered.}
  \label{fig:novel_data_metadata}
\end{figure}

\begin{figure}[h]
  \centering
  \includegraphics[width=\textwidth]{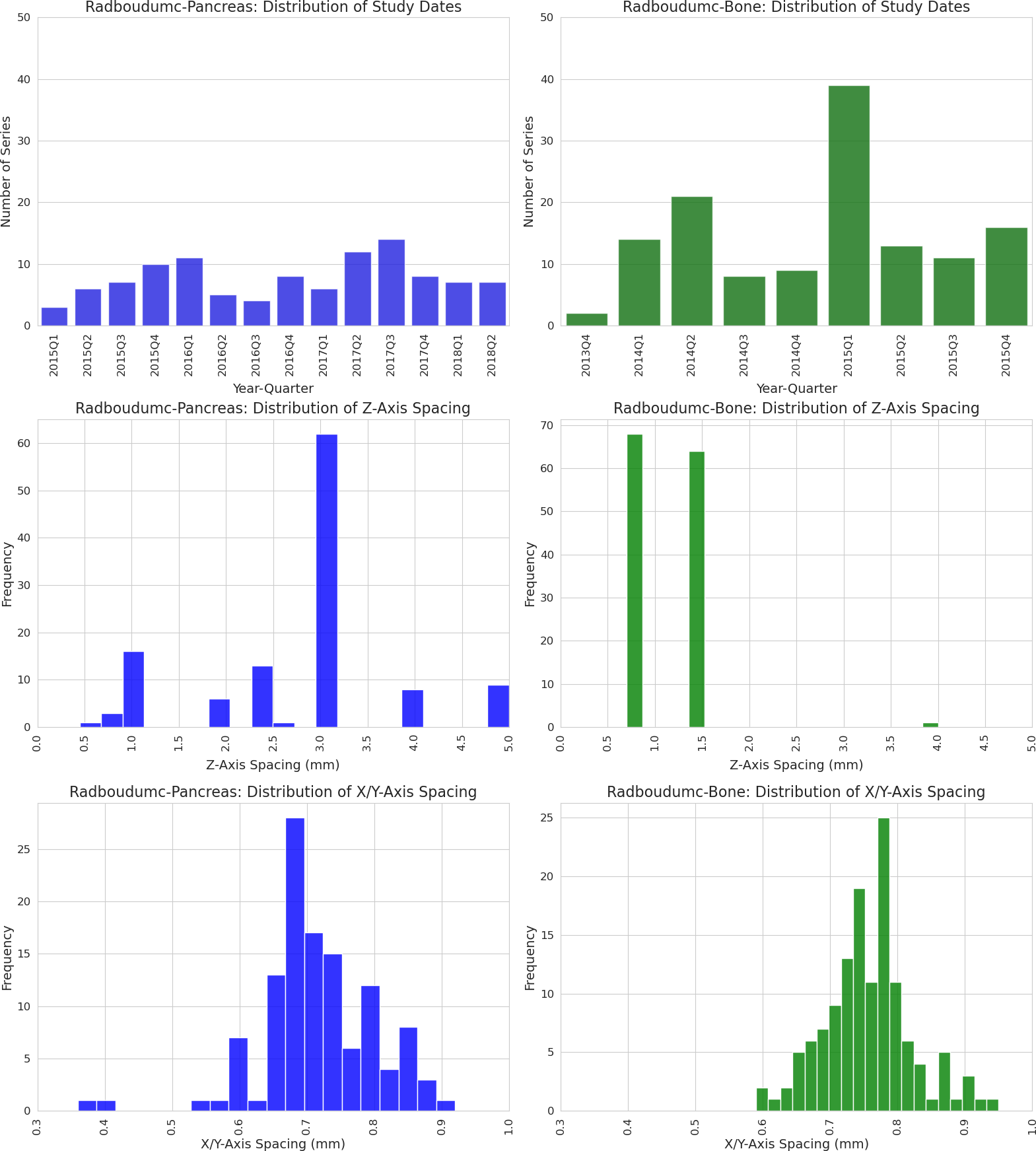}
  \caption{Study date and scan spacing distributions for the series included in the two novel training datasets.}
  \label{fig:additional_metadata}
\end{figure}

\begin{figure}[h]
  \centering
  \includegraphics[width=\textwidth]{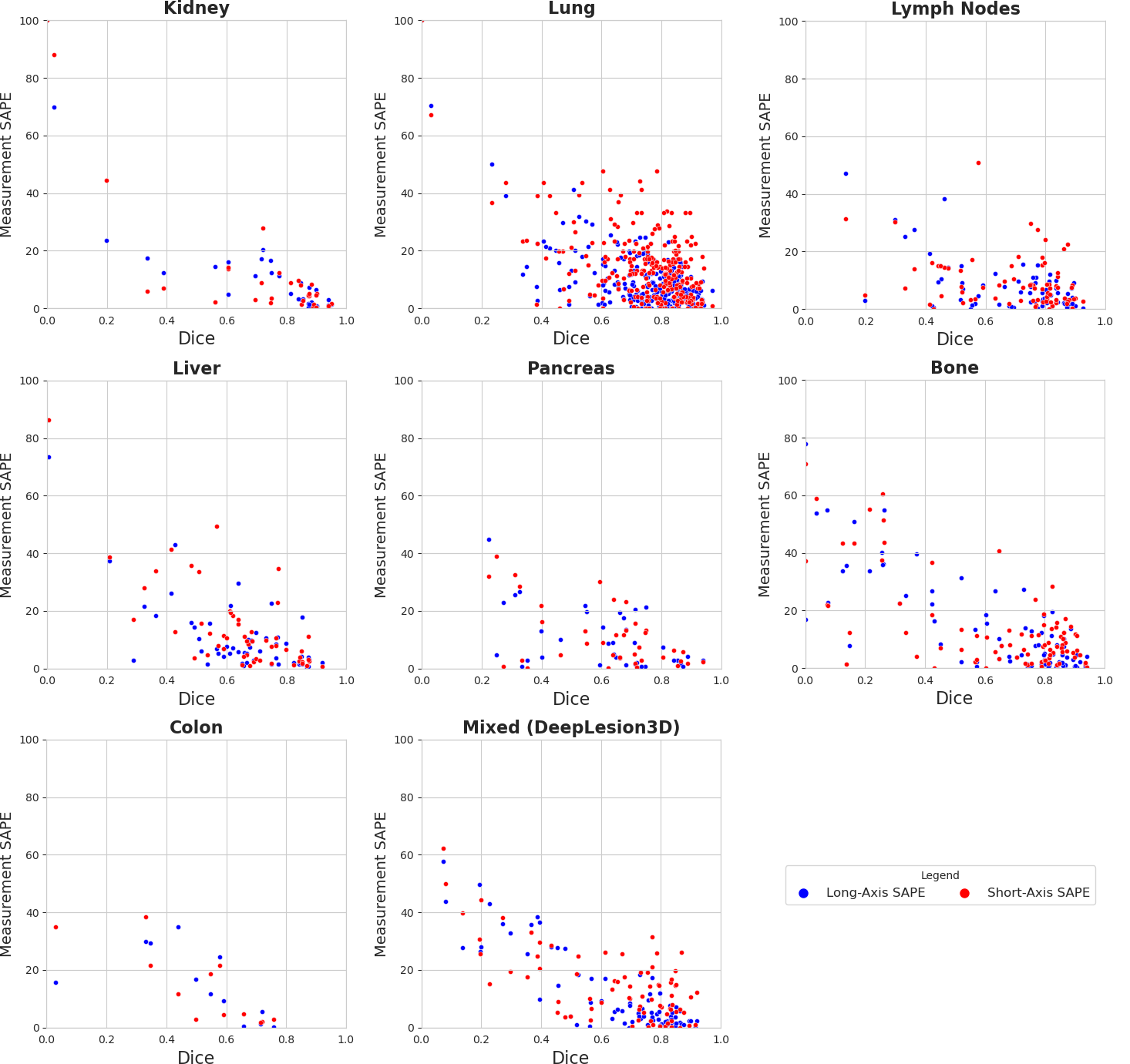}
  \caption{Plots of the Dice score vs the long- and short-axis measurement error for the baseline model on the different lesion types in the held-out training data. SAPE = Symmetric Absolute Percentage Error}
  \label{fig:error_plots_held_out}
\end{figure}

\begin{figure}[h]
  \centering
  \includegraphics[width=\textwidth]{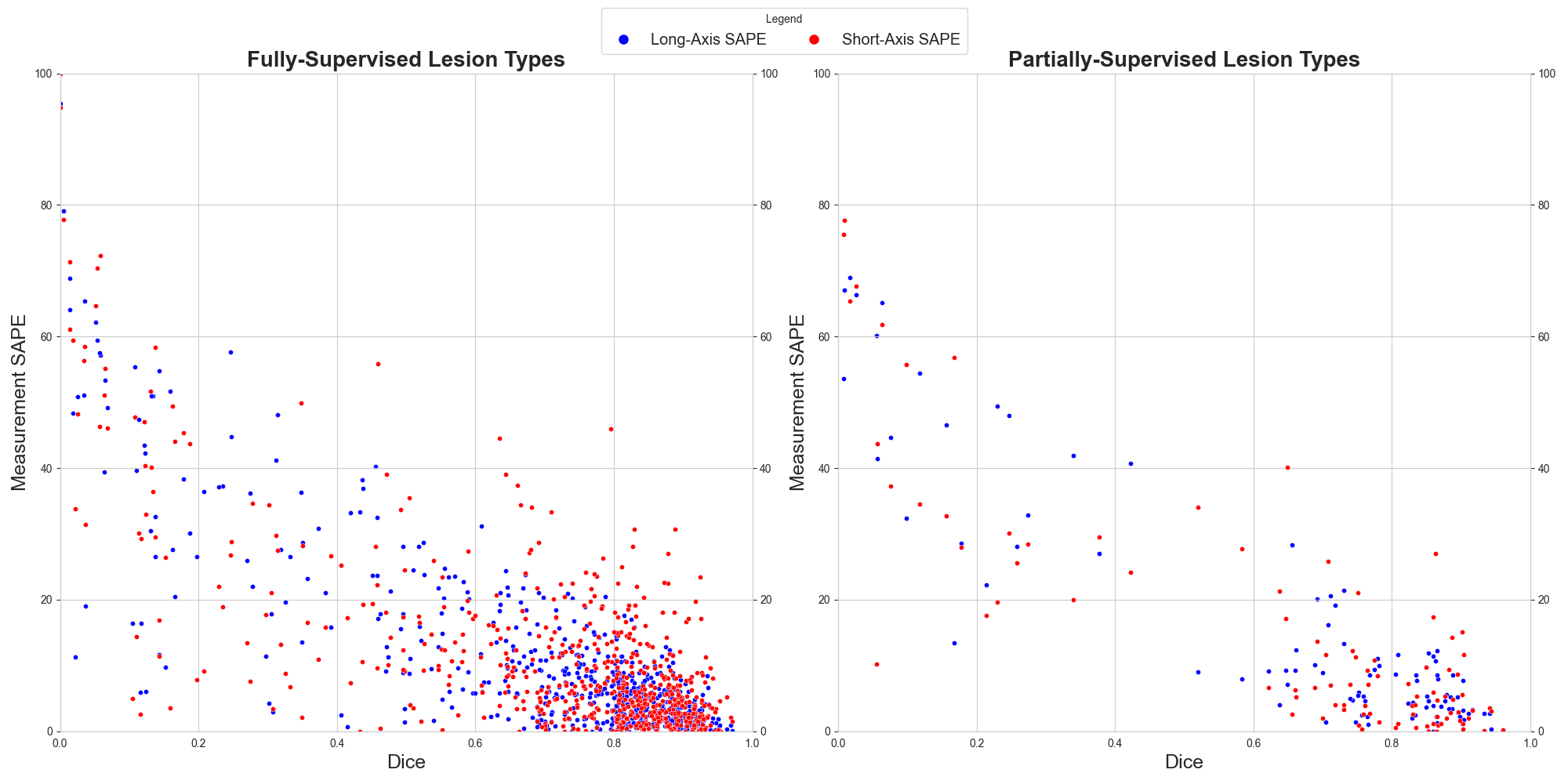}
  \caption{Plots of the Dice score vs the long- and short-axis measurement error for the baseline model on the test data, split on lesion types seen in the fully-annotated data versus those in the partially annotated data. SAPE = Symmetric Absolute Percentage Error}
  \label{fig:error_plots_test}
\end{figure}

\subsection{Additional Results} 
\label{sec:cds:appendixB1}
\begin{figure}[h]
  \centering
  \includegraphics[width=\textwidth]{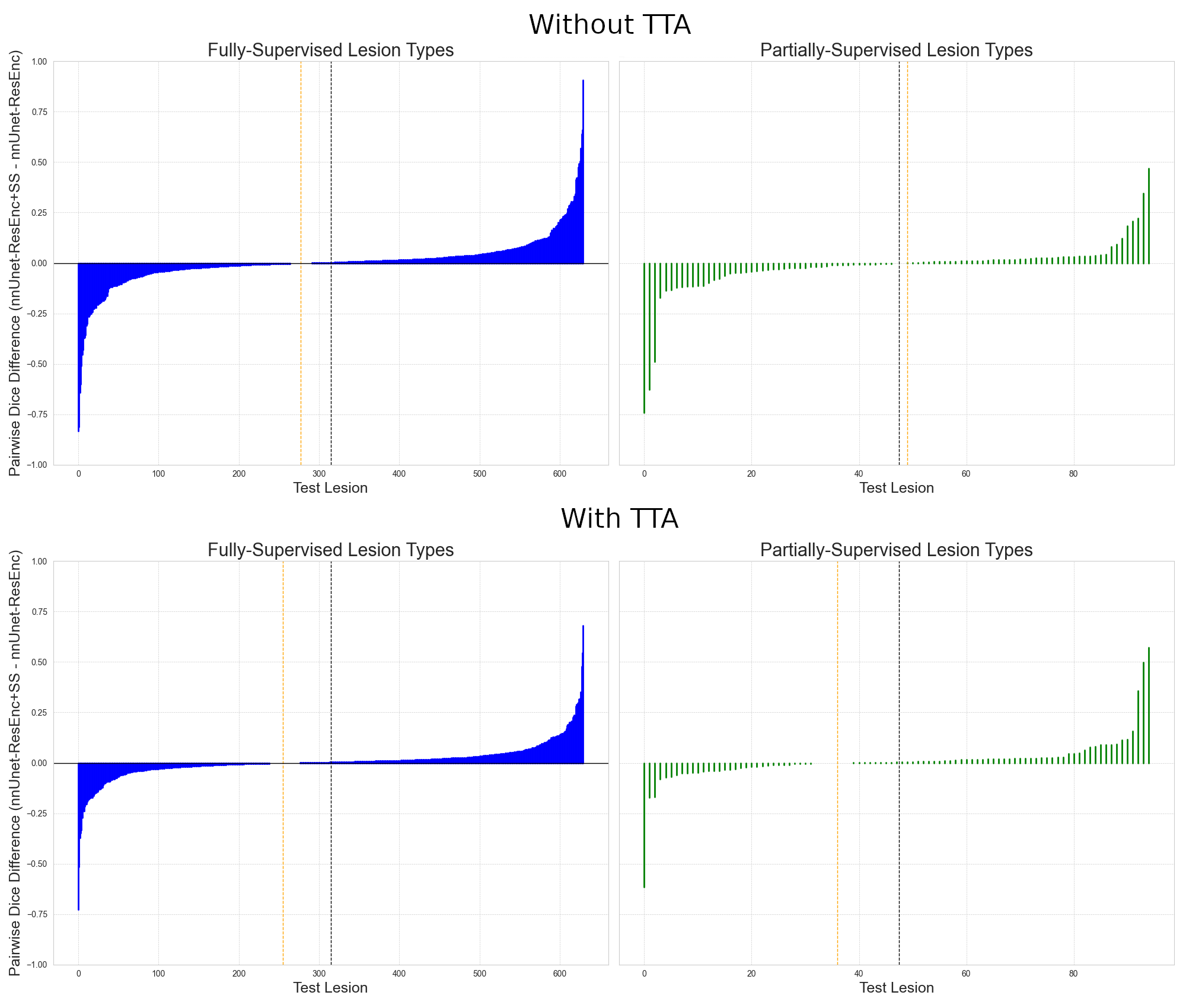}
  \caption{Plots of the sorted pairwise difference in Dice score between the up-scaled residual encoder nnUnet trained with and without semi-supervision. The left graphs contain the lesion types in the test set covered by the fully-annotated training data, the right graphs contain the scores for the lesion types only present in the partially-annotated data. A negative score indicates the segmentation performance of the fully-supervised nnUnet was better for that case, a positive score indicates the semi-supervised nnUnet scored higher for that case. The black vertical line indicates 50\% of the lesions, the orange line denotes where the score changed from negative to positive. TTA = Test Time Augmentations.}
  \label{fig:pairwise_test}
\end{figure}

% With TTA
By default, the nnUnet framework applies Test Time Augmentation (TTA) during inference by mirroring the input volume across all axes. This means that for each VOI, the model needs to predict for eight different views, by mirroring across one, two, or all three axes. Usually, this improves performance, but it comes at a significant runtime cost. Due to the strict time limit imposed by the challenge design, we opted not to use this feature for our baseline model. Without TTA, our models can run inference on a VOI in under 2 seconds using a NVIDIA RTX2080 TI GPU. However, in scenarios that do not require immediate predictions, TTA can easily be incorporated. Table \ref{tab:results_with_tta} shows the impact TTA has on our models. Figure \ref{fig:pairwise_test} shows how with TTA the semi-supervised model improves relative to the residual encoder nnUnet when comparing the pairwise Dice score per case in the test set. Without TTA, we can see that the semi-supervised model already achieves slightly more positive scores than the standard nnUnet (the orange line indicating the transition point from negative to positive scores, with the black line indicating half of the data). With TTA, this effect becomes more pronounced, especially for those lesion types not covered in the fully annotated lesions. We can also see how there are fewer cases with large negative scores when using TTA.

\begin{table}[h]
\resizebox{\textwidth}{!}{
\begin{tabular}{ccccccccc}
 &  &  &  &  & Dice &  &  &  \\ \hline
 & \multicolumn{1}{c|}{} & \textbf{Kidney ($N=33$)} & \textbf{Lung ($N=267$)} & \textbf{Lymph-Node ($N=77$)} & \textbf{Bone ($N=86$)} & \textbf{Liver ($N=54$)} & \textbf{Pancreas ($N=38$)} & \textbf{Colon ($N=12$)} \\ \hline
 Held-Out & \multicolumn{1}{l|}{nnUnet} & $0.621 \pm 0.258$ & \underline{$0.761 \pm 0.138$} & \underline{$0.709 \pm 0.165$} & $0.661 \pm 0.284$ & \underline{$0.651 \pm 0.167$} & $0.579 \pm 0.221$ & $0.505 \pm 0.245$ \\
 Training Data  & \multicolumn{1}{l|}{nnUnet-ResEnc} & $0.722 \pm 0.234$ & $0.757 \pm 0.144$ & $0.700 \pm 0.176$ & $0.656 \pm 0.262$ & $0.650 \pm 0.173$ & $0.624 \pm 0.205$ & $0.471 \pm 0.222$ \\
 & \multicolumn{1}{l|}{nnUnet-ResEnc+SS} & \underline{$0.768 \pm 0.209$} & $0.758 \pm 0.138$ & $0.699 \pm 0.179$ & \underline{$0.677 \pm 0.239$} & $0.646 \pm 0.175$ & \underline{$0.636 \pm 0.175$} & \underline{$0.548 \pm 0.210$} \\
 &  &  &  &  &  &  &  &  \\ \hline
 & \multicolumn{1}{c|}{} & \textbf{Kidney}* & \textbf{Lung} & \textbf{Lymph-Node} & \textbf{Bone}* & \textbf{Liver} & \textbf{Pancreas}* & \textbf{Colon}* \\ \hline
 \multirow{2}{*}{Test Set} & \multicolumn{1}{l|}{nnUnet} & $0.586 \pm 0.215$ & $0.719 \pm 0.252$ & $0.665 \pm 0.230$ & $0.308 \pm 0.233$ & $0.637 \pm 0.256$ & $0.551 \pm 0.149$ & $0.476 \pm 0.225$ \\
 & \multicolumn{1}{l|}{nnUnet-ResEnc} & \underline{$0.695 \pm 0.197$} & $0.760 \pm 0.215$ & $0.697 \pm 0.213$ & $0.398 \pm 0.264 $ & $0.706 \pm 0.250$ & $0.772 \pm 0.101$ & $0.410 \pm 0.052$ \\
 & \multicolumn{1}{l|}{nnUnet-ResEnc+SS} & $0.670 \pm 0.182$ & \underline{$0.764 \pm 0.214$} & \underline{$0.702 \pm 0.230$} & \underline{$0.412 \pm 0.289$} & \underline{$0.726 \pm 0.228$} & \underline{$0.774 \pm 0.098$} & \underline{$0.510 \pm 0.213$} \\
 &  &  &  &  &  &  &  &  \\ \cline{1-5}
 & \multicolumn{1}{c|}{} & \textbf{Full Test Set ($N=725$)} & \textbf{FSUP ($N=630$)} & \textbf{PSUP ($N=95$)} &  &  &  &  \\ \cline{1-5}
 \multirow{2}{*}{Test Set} & \multicolumn{1}{l|}{nnUnet} & $0.665 \pm 0.253$ & $0.671 \pm 0.248$ & $0.626 \pm 0.286$ &  &  &  &  \\
 & \multicolumn{1}{l|}{nnUnet-ResEnc} & $0.707 \pm 0.235$ & $0.715 \pm 0.223$ & $0.649 \pm 0.288$ &  &  &  &  \\
 & \multicolumn{1}{l|}{nnUnet-ResEnc+SS} & \underline{$0.715 \pm 0.237$} & \underline{$0.723 \pm 0.228$} & \underline{$0.661 \pm 0.287$} &  &  &  & 
\end{tabular}
}
\caption{Segmentation performance comparison with test time augmentation on the 10\% held-out training data per lesion type and the test set. For the individual lesion types in the test set, * indicates there were $\leq$ 20 lesions of this type in the test set. The exact distribution of lesion types is not provided to participants. FSUP = fully supervised lesions types (i.e. Kidney - Colon). PSUP = lesion types present in the partially supervised training data.}
\label{tab:results_with_tta}
\end{table}

\begin{figure}[h]
  \centering
  \includegraphics[width=\textwidth]{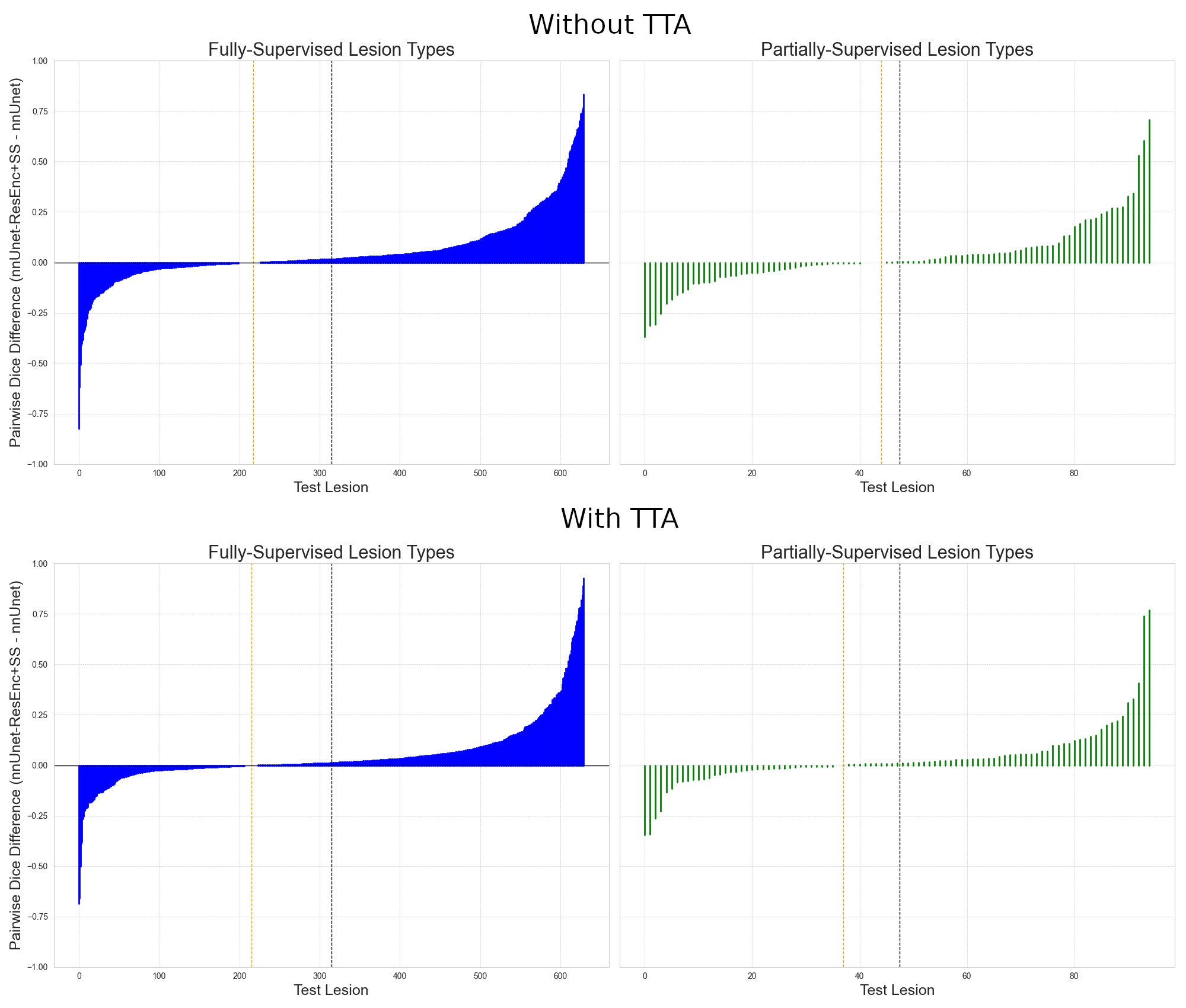}
  \caption{Plots of the sorted pairwise difference in Dice score between the up-scaled residual encoder nnUnet trained with semi-supervision and the self-configured nnUnet. The left graphs contain the lesion types in the test set covered by the fully-annotated training data, the right graphs contain the scores for the lesion types only present in the partially-annotated data. A negative score indicates the segmentation performance of the regular nnUnet was better for that case, a positive score indicates the semi-supervised nnUnet scored higher for that case. The black vertical line indicates 50\% of the lesions, the orange line denotes where the score changed from negative to positive. TTA = Test Time Augmentations.}
  \label{fig:pairwise_test_regular}
\end{figure}

\subsection{Model Plans}
\label{sec:cds:appendixB2}
To use the nnUnet framework for our task, we needed to make two minor adjustments. First, we integrated a novel resampling function that bypasses the alteration of image sizes based on the dataset fingerprint. Standardizing the scan resolution to the parameters extracted by the framework, which looks at the median spacing, often resulted in a loss of important details on high-resolution scans. The nnUnet framework was not designed with the large diversity of resolutions of our combined training datasets in mind. Resampling is also time-intensive, and since our model needs to process each VOI in under 3 seconds, eliminating it altogether allowed us to significantly accelerate inference. Secondly, we directly altered the plans file to prevent the model from patching the VOIs. This step is needed to ensure that the model retains the information that the lesion to be segmented is located at the center of the VOI. This is crucial since VOIs may contain multiple lesions in close proximity, and only one needs to be segmented. The hyperparameters of the baseline model can be found in Table \ref{tab:hyperparameters}, along with the intensity normalization values for the data.

\begin{table}[h]
\centering
\scriptsize
\begin{tabular}{lllll}
\cline{1-2} \cline{4-5}
\textbf{\textbf{Parameter}} & \textbf{\textbf{Setting}} &  & \textbf{\begin{tabular}[c]{@{}l@{}}Intensity Properties\end{tabular}} & \textbf{Value (HU)} \\ \cline{1-2} \cline{4-5} 
unet\_class\_name & ResidualEncoderUNet &  & min & -2048 \\
normalization\_schemes & CTNormalization &  & max & 3071 \\
use\_mask\_for\_norm & false &  & median & 51 \\
batch\_size & 3 &  & mean & 21.7 \\
unet\_base\_num\_features & 32 &  & percentile 0.5\% & -910 \\
unet\_max\_num\_features & 384 &  & percentile 99.5\% & 1672 \\
num\_pool\_per\_axis & [5,6,6] &  & std & 331.6 \\ \cline{4-5} 
pool\_op\_kernel\_sizes & [[1,1,1], [2,2,2]*5, [1,2,2]] &  &  &  \\
conv\_kernel\_sizes & [[3,3,3]*7] &  &  &  \\
n\_conv\_per\_stage\_encoder & [1,3,4,6,6,6,6] &  &  &  \\
n\_conv\_per\_stage\_decoder & [1,1,1,1,1,1] &  &  &  \\
resampling & false &  &  &  \\
batch\_dice & false &  &  &  \\
epochs & 1000 &  &  &  \\
starting lr & 1e-2 &  &  &  \\ \cline{1-2}
\end{tabular}
\caption{Hyperparameters of the baseline model (nnUnet-ResEnc+SS) and the intensity properties used for data normalization, calculated from the pretraining data.}
\label{tab:hyperparameters}
\end{table}

\subsection{Held-Out Training Data Split}
\label{sec:cds:appendixB3}

Table \ref{tab:data_split} contains the identifiers of the patients from the training data that were used to evaluate our models during development. They represent a random, patient-level, 10\% split of each dataset making up the fully-annotated training data. For the evaluations per lesion type we pooled the lesions from the different related datasets, e.g. MDSC-Lung, LIDC-IDRI and LNDb for lung lesions.

\begin{table}[h]
\centering
\scriptsize
\begin{tabular}{|l|l|l|p{6.75cm}|}
\hline
\textbf{Dataset} & \textbf{No. Patients} & \textbf{No. Lesions} & \textbf{Patient IDs} \\ \hline
MDSC-Lung & 6 & 10 & 14, 27, 43, 64, 78, 84 \\ \hline
MDSC-pancreas & 27 & 27 & 5, 16, 19, 25, 35, 40, 42, 67, 109, 135, 140, 165, 183, 203, 204, 213, 256, 279, 301, 302, 323, 329, 348, 354, 357, 391, 402 \\ \hline
MDSC-Colon & 12 & 12 & 7, 26, 78, 89, 99, 106, 126, 129, 155, 161, 164, 192 \\ \hline
NIH-LN-MED & 8 & 38 & 1, 30, 31, 33, 48, 65, 77, 86 \\ \hline
NIH-LN-ABD & 8 & 40 & 7, 19, 28, 35, 46, 54, 74, 76 \\ \hline
LIDC-IDRI & 72 & 207 & 2, 3, 30, 38, 40, 49, 50, 70, 71, 97, 103, 124, 129, 132, 133, 149, 163, 167, 182, 194, 198, 200, 213, 223, 228, 230, 246, 252, 270, 285, 286, 289, 297, 305, 309, 312, 313, 321, 329, 330, 339, 340, 346, 363, 382, 389, 397, 407, 429, 431, 436, 460, 494, 526, 528, 529, 531, 536, 538, 616, 629, 648, 649, 663, 666, 674, 684, 688, 720, 738, 741, 750 \\ \hline
LNDb & 19 & 53 & 5, 20, 25, 28, 34, 67, 68, 74, 80, 84, 97, 110, 144, 157, 180, 209, 229, 258, 304 \\ \hline
LiTS & 9 & 54 & 2, 4, 16, 19, 23, 39, 67, 80, 109 \\ \hline
KiTS21 & 30 & 33 & 3, 36, 71, 82, 84, 109, 117, 120, 121, 140, 146, 153, 165, 170, 173, 185, 187, 195, 196, 205, 216, 221, 222, 238, 239, 244, 249, 273, 275, 288 \\ \hline
Radboudumc-Pancreas & 11 & 11 & 47, 266, 268, 313, 336, 1014, 2002, 2003, 2008, 2015, 2072 \\ \hline
Radboudumc-Bone & 14 & 88 & 30, 203, 209, 233, 281, 289, 293, 294, 309, 346, 578, 797, 1159, 1188 \\ \hline
DeepLesion3D & 49 & 78 & 102, 103, 116, 204, 218, 277, 383, 404, 407, 453, 523, 550, 569, 653, 677, 760, 858, 904, 985, 1086, 1118, 1394, 1409, 1564, 1969, 2123, 2136, 2177, 2249, 2257, 2274, 2328, 2350, 2480, 2521, 2694, 2713, 2779, 2900, 2958, 3040, 3171, 3191, 3327, 3793, 3880, 4035, 4049, 4083 \\ \hline
\end{tabular}
\caption{List of the patient IDs per dataset in the held-out training data }
\label{tab:data_split}
\end{table}
\end{document}